\documentclass[a4paper,10pt]{article}
\date{}
\begin{document}

\title{ Can Smoluchowski equation account for gelation transition? }
\author{{\small M. K. Hassan$^{1,2}$ and J. Kurths$^1$} \\ {\small $~^1$ University of Potsdam,
Department of Physics, Am Neuen Palais, D-14415, Potsdam, Germany} \\
 {\small $~^2$  University of Dhaka, Department of Physics, Theoretical Physics
Division, Dhaka 1000, Bangladesh}}

\maketitle

\begin{abstract} 

\noindent
We revisit the scaling theory of the Smoluchowski equation with special emphasis
on the dimensional analysis to derive the scaling ansatz and to give an insightful foundation to it. 
It has long been argued that the homogeneity exponent $\lambda$ of the 
aggregation kernel divides the aggregation process into two regimes (i) $\lambda\leq 1$ nongelling
and (ii) $\lambda>1$ gelling. However, our findings contradict with this result. In particular, 
we find that the Smoluchowski equation is valid if and only if $\lambda<1$. We show that beyond this limit 
i.e. at $\lambda\geq 1$, it breaks down and hence it fails to describe a gelation transition.     
This also happens to be accompanied by violation of scaling.

\end{abstract}  

\vspace{3mm}

\noindent
PACS number(s): 05.20.Dd,02.50.-r,05.40-y

\vspace{2mm}

\noindent
The kinetics of irreversible and sequential aggregation of particles  
occurs in a variety of physical processes and it is of wide interest in physics, 
chemistry, biology and in many other disciplines 
of science and technology \cite{kn.has1,kn.has2,kn.has3}.  
Due to its cross-disciplinary importance, statistical physics has offered a
number of models: diffusion limited aggregation (DLA) \cite{kn.has4}, percolation \cite{kn.has5}, 
diffusion limited cluster cluster aggregation (DLCA), ballistic aggregation (BA) 
\cite{kn.has6} etc., to name just a few, and these are systems far from equilibrium.
There  hardly exists any systematic standard theoretical framework for describing the 
systems out of equilibrium, which is 
indeed in sharp contrast to its equilibrium counterpart. 
However, stochastic processes seem to some extent to rescue this shortcoming and 
appear to capture a wide class of non-equilibrium phenomena \cite{kn.frag}. 
The dynamics of these processes often evolves in
time following some conservation rules and can be expressed in the form of a master equation
that constitutes a relation between the spatial and its temporal variables.

\noindent
As far as kinetics of cluster-cluster aggregation is concerned, much of the 
theoretical understanding is provided  by the rate 
equation approach proposed by von Smoluchowski \cite{kn.has7} almost a century ago which reads as   
\begin{equation}
\partial_t \psi(x,t)  = -  a(x,t)\psi(x,t)+{{1}\over{2}}\int_0^x dy K(y,x-y)\psi(y,t)\psi(x-y,t). 
\end{equation}
This is a one-dimensional mean-field model, where $\psi(x,t)$ is the  
concentration of particles of size $x$ at time $t$. We note for the first time that the kernel
$K(x,y)$ itself is not the aggregation rate but rather it is the $a(x,t)=\int_0^\infty K(x,y)\psi(y,t)dy$
which is the rate at which particles aggregate and hence it has to be obtained self-consistently.
Equation $(1)$ is studied for a large class of 
aggregation kernels that satisfy the power-law condition
$K(bx,by)=b^\lambda K(x,y)$,
which has physical importance too. It ensures that the aggregating particles are homogeneously distributed, i.e. 
that the system has a perfect mixing of particles with continuous distribution of sizes
and hence we call $\lambda$ the homogeneity exponent. This $\lambda$ value appears to 
play a significant role in controlling the dynamics of the
process. However, to the best of our knowledge, the limit 
on the $\lambda$ value up to which Eq. $(1)$ 
is valid,  or if such limits exist at all, has never been studied.

\noindent
In this article we, therefore, revisit the scaling theory of Smoluchowski 
equation with a special emphasis on the dimensional analysis in an 
attempt to obtain a deep insight into the dynamics of the process and find the  
limits on the $\lambda$ value upto which the Smoluchowski equation is valid. We get several results that are 
in contradiction with existing predictions, 
especially those referring to the gelling-nongelling transition. It has 
long been argued that the homogeneity
exponent $\lambda$ of the aggregation kernel divides the aggregation process 
into two regimes \cite{kn.has8,kn.has14}. First, the nongelling model described by $\lambda\leq 1$,  
where the Smoluchowski equation admits scaling and obeys mass conservation. Second,  
the gelling model at $\lambda>1$, where it is believed that the gelation transition 
occurs after a certain time $t>t_{cr}$ and it 
is accompanied by the violation of scaling and mass conservation. However, it is claimed
that at $t<t_{cr}$ the equation admits scaling and obeys mass conservation. 
In this letter, we question the validity of the equation in the gelling regime and show that the 
Smoluchowski equation is valid if and only if $\lambda<1$.   
Beyond this bound, including $\lambda=1$, we show that Eq. (1) violates 
the basic principle of aggregation for the whole range of the time axis.  
To further support our argument, we consider the best known gelling model 
and show that the results obtained from this model are either unphysical or self-contradictory.
Note that the sol-gel transition is in 
fact a kinetic phase transition that is characterized
by a singularity of typical quantity, such as enthalpy accompanied by a sharp rise in 
viscosity and appearance of non-zero shear modulus
at the critical point. It is important to emphasize that Smoluchowski's equation
is a one dimensional mean-field model. It does not take into account the detailed
nature of cluster-cluster interactions, their spatial correlation and does not include any 
thermodynamical or mechanical
behaviour, instead it only describes some average behaviour of the underlying mechanism through the coagulation kernel.
Therefore, we argue that like many other one dimensional mean-field models
in statistical physics it cannot account for phase transition of any order and  sol-gel transition in particular.

\vspace{3mm}

\noindent
We start with appreciating that Eq. (1) describes a stochastic process; the particle size $x$ and the time 
$t$ are the only two governing parameters that can completely specify the governed 
parameter $\psi(x,t)$. However, it is 
very instructive to note that $x$ and $t$ are    
intertwined via the dimensional consistency of Eq. $(1)$. 
Of course, the exact relation will depend on $a(x,t)$ through the precise choice of $K(x,y)$.
As a consequence of this, either of the two can be taken to be an independent parameter when the 
other governing parameter and the governed parameter $\psi(x,t)$ is expressible in terms of this. 
For example, if $t$ is chosen to be the independent
parameter, then using the fact that the dimension of a physical quantity is always
expressed as power monomial, we can define the dimensionless quantity
$\xi={{x}\over{t^z}}$ \cite{kn.barenblatt}. Then, it is obvious that the dimension of the governed parameter
$\psi(x,t)$ can also be written only in terms of $t$, and hence
we can define another dimensionless quantity
$\Pi \equiv {{\psi(x,t)}\over{t^{\alpha}}} \sim t^{-\alpha} \psi(\xi t^z,t)\equiv
F(\xi,t)$.
Since $\xi$ and $\Pi$ are both dimensionless quantities, upon transition from one system of 
units of measurement to another, inside a given class, their numerical values must 
remain unchanged. In other words, we can pass to a system of unit of measurement where $t$ 
is changed by an arbitrary factor and, upon such a transition, the function $F$ or $\Pi$ 
must remain unchanged, i.e.
${{\partial F}\over{\partial t}}=0$.
This implies that the quantity $\Pi$ is independent of $t$ and can be completely 
expressed in terms of $\xi$ alone. Thus we can define  
$\Pi=\phi(\xi)$ to enable us to write the temporal scaling ansatz
$\psi(x,t)\approx t^\alpha \phi({{x}\over{t^z}})$. Note that $z$ and $\alpha$ can only take values
for which $t^z$ and $t^{\alpha}$ bear the dimension of $x$ and $\psi$ respectively.
Had we chosen $x$ to be the independent 
parameter, a similar argument would lead us to write the spatial scaling ansatz
$\psi(x,t)\approx x^{-\theta}\Phi({{t}\over{x^{-\nu}}})$, where $x^{-\nu}$ and  $x^{-\theta}$ must
bear the dimension of $t$ and $\psi(x,t)$ respectively.
The existence of scaling or a self-similar solution actually means that we can choose the self-similar
coordinates ${{\psi}\over{t^\alpha}}$ (or ${{\psi}\over{x^{- \theta}}}$) and
$x/t^z$ (or $t/x^{-\nu}$) such that their plots for any initial condition collapse
into one single curve, which is typically known as {\it data collapse}
formalism \cite{kn.barenblatt,kn.stanley}.

\noindent
To find the scaling exponents and for further use, we define the $n$th moment 
 of $\psi(x,t)$ as $M_n(t)=\int_0^\infty dx x^n \psi(x,t)$ with $n\geq 0$.   
We then get a rate equation for $M_n(t)$ upon multiplying both sides of 
Eq. $(1)$ by $x^n$ and integrate over the whole range of $x$; this yields  
\begin{equation}
{{dM_n(t)}\over{dt}}={{1}\over{2}}\int_0^\infty \int_0^\infty dx 
dy[(x+y)^n-x^n-y^n]K(x,y)\psi(x,t)\psi(y,t).
\end{equation}
This immediately reveals that $M_1(t)$, the mass or size of the system is a
conserved quantity, since Eq. $(1)$ describes aggregation of particles in a closed system.  
This conserved quantity in fact is the intrinsic agent responsible for fixing the 
numerical value of the mass exponent  $\alpha=-2z$ (or $\theta=2$). 
On the other hand, the $z$ (or $\nu$) value is fixed by the 
aggregation rate $a(x,t)$ and hence it is called the kinetic exponent.
It is quite easy to realize that a general explicit solution for $M_n(t)$ by
solving a non-linear equation, like Eq. $(2)$, can be extremely difficult for any 
choice of $K(x,y)$. Therefore, we will resort to 
an indirect way of obtaining it, namely the dimensional analysis and find the asymptotic
behaviour of $M_n(t)$ instead.  
Note that the kinetic exponent $z$ can 
only be calculated if the aggregation kernel is specified with respect to its argument.
For example, we select $K(x,y)=(xy)^\beta$ (i.e. $\lambda=2\beta$) and call it the product model.
For this choice of aggregation kernel, we substitute the temporal scaling ansatz into the rate
equation $(1)$ and get 
\begin{equation}
-zt^{z-2\beta z-1}={{-\xi^\beta\phi(\xi)\phi_\beta+{{1}\over{2}}\int_0^\xi d\eta 
\eta^\beta (\xi-\eta)^\beta 
\phi(\eta)\phi(\xi-\eta)}\over{2\phi(\xi)+\xi\phi^\prime(\xi)}}.
\end{equation}
Since the left hand side of the above equation is a function of $t$ only, whereas the right 
hand side is independent of $t$ and a dimensionless quantity, the exponent of $t$ must be 
equal to zero to give $z={{1}\over{1-2\beta}}$. This means $t^{{{1}\over{1-2\beta}}}$ bears 
the dimension of the quantity $x$, in other words 
$x^{-(2\beta-1)}$ contains the dimension of $t$  i.e. $\nu=2\beta-1$. For this product 
model the quantity $a(x,t)= x^\beta M_\beta(t)$, 
being the rate of aggregation, must bear the dimension inverse of time. Combining all this dimensional consistency 
we get $M_\beta \sim x^{\beta-1}$ or $a(x,t)\sim x^{2\beta-1}$ and a trivial change 
in variable gives $M_n(t)\sim t^{{{n-1}\over{1-2\beta}}}$.

\noindent
To further generalize our argument, we consider $K(x,y)=(x+y)^\gamma$ and $K(x,y)=x^\gamma+y^\gamma$ 
(i.e. $\lambda=\gamma$) and call both the sum model 
for which the aggregation rates $a(x,t)$ are $\sum_{r=0}^\gamma 
\left( \begin{array}{c}\gamma \\ r\end{array}\right )x^rM_{\gamma-r}$ 
and $(x^\gamma M_0(t)+ M_\gamma(t))$ respectively.
Substituting the scaling ansatz into Eq. $(1)$, we obtain  
an equation similar to Eq. $(3)$ for both models and both give
$z={{1}\over{1-\gamma}}$. Doing further dimensional analysis gives $a(x,t) \sim x^{\gamma-1}$ and
$M_n(t) \sim t^{{{n-1}\over{1-\gamma}}}$. 
Comparing the results 
of all the three models, we can single out one exponent $\lambda$ 
and generalize the expression for the moments $M_n(t)$ and the kinetic exponent $z$.
That is, for any homogeneous kernel 
with a homogeneity exponent $\lambda$ we postulate $z={{1}\over{1-\lambda}}$ and 
$M_n(t)\sim t^{(n-1)z}$. 
We now define the simple and weighted mean cluster size as $s(t)=M_1(t)/M_0(t)$ and $w(t)=M_2(t)/M_1(t)$ 
respectively. However, both the definitions give the same result $s(t)=w(t)=t^z$ and this again confirms
that $t^z$ bears the dimension of the particle size $x$.  
Note that the kinetic exponent $z$ suffers a singularity at $\lambda=1$, otherwise $z>0$ if  $\lambda<1$ and
$z<0$ if $\lambda>1$. 
Therefore, it is obvious that the behaviour of $s(t)$ should also change with the change of the $\lambda$ value.
The expression for $s(t)$ in fact tells us how the mean cluster size should 
evolve in the system. 
The basic principle of the aggregation process in a closed system  
is that the number of particles present in the system must be a decreasing function of time
and therefore $s(t)$ should be an ever growing quantity. 
Clearly, this sets a physical constraint
on the choice of the $\lambda$ value in determing the bound beyond which $s(t)$ violates the basic principle.
This is what happens when $\lambda>1$ and $s(t)$ becomes a decreasing quantity in time.

\noindent
Before discussing the so called 
gelling regime $\lambda>1$, we shall first show that at $\lambda=1$ the scaling is violated, which contradicts with
the known results \cite{kn.has10}.
In this case $z$ suffers a singularity and gives $a(x,t)=1$. 
This means that the relation between the two 
governing parameters $x$ and $t$ no longer exists 
and therefore $x$ cannot be expressed in terms
of $t$ and vice versa.  
In this situation, on the one hand, the power monomial character of the system is lost and, 
on the other, one can no longer define the 
self-similar coordinates which is indispensible for any process to show scaling; this proves 
that at $\lambda=1$ the scaling is violated \cite{kn.barenblatt,kn.stanley}.
This can be further supported by the explicit known solution for $\gamma=1$ (i.e. $\lambda=1$) 
for which one obtains $s(t)=e^t$ \cite{kn.has10, kn.has11}. This clearly shows that 
both $s(t)$ and $t$ have lost their dimensional character.
We argue that to admit scaling (i) $s(t)$ must show a power-law behaviour and (ii)
 owing to the nature of the process in question (kinetics of aggregation), the exponent 
has to be a positive and a non-zero finite quantity \cite{kn.barenblatt}. Therefore, at $\lambda=1$
not only the scaling is violated but also Eq. (1) fails to describe a physically meaningful aggregation process 
since the parameters that govern the system lose their dimensional (stochastic) character.  

\vspace{3mm}

\noindent
We now show that the scaling is also violated for $\lambda>1$, irrespective of whether we are
below or above $t_{cr}$. First, we argue that the solution of Eq. (1) must be of a stretched exponential
decay type so that all the moments exist and show a power-law behaviour. The justification of this rests on the 
loss term which dictates that $\psi(x,t)$ must decay exponentially.  
The argument of the exponential term in the solution must be $\xi \sim a(x,t)t$. That is,
it can be either $\xi\sim t/x^{1-\lambda}$ 
 or $\xi\sim x/t^{1/(1-\lambda)}$ since $a(x,t)\sim x^{\lambda-1}$.
The important point here is that, this exponential decay term must be present even in the long time behaviour
of the solution (scaling solution) so that all the moments exist and show a power-law behaviour.
Otherwise, the Eq. (1) itself becomes invalid since 
the aggregation rate $a(x,t)$ itself is a function of $M_n(t)$, where $n$
depends on the specific choice for the kernel. The scaling solutions of the 
aggregation process are essentially the solutions in the long-time ($t \longrightarrow \infty$)
and large-size ($x \longrightarrow
\infty$) limit so that in this regime Eq. (1) reduces itself to an ordinary differential equation for the scaling
function $\phi(\xi)$ \cite{kn.has15}. That is, the two governing 
parameters $x$ and $t$ must combine together to form 
a dimensionless quantity $\xi$ in such a way that in the scaling regime $\xi$ may stay finite.
This is possible if and only if $\xi=x/t^z$ or $\xi=t/x^{1/z}$ with $z>0$ or $\lambda<1$. 
It is important to mention here the close connection between the rate
equation approach of fragmentation \cite{kn.has15} and the aggregation process described by  
Eq. (1). In both cases, the mean cluster size and the concentration 
have the same functional relation with time i.e. $s(t)\sim t^z$ and $\psi(x,t)\sim t^{-2z}\phi(x/t^z)$ respectively.
The mean cluster $s(t)$ must bear the dimenion of $x$ while the concentration $\psi(x,t)$ must bear 
the dimension of $M_0(t)/s(t)\sim t^{-2z}$, since it is defined as the number of particles per unit length, 
irrespective of whether $s(t)$ and $\psi(x,t)$ describe the aggregation or
the fragmentation process. 
The two opposing phenomena keep the signature of their respective processes through their $z$ value i.e.
$z<0$ for the fragmentation and $z>0$ for the aggregation process. All the existing solutions in both  
phenomena agree with this provided we are in the valid regime.
The scaling solutions of the fragmentation equation are on the other hand solutions in the long-time  
($t \longrightarrow \infty$) and the small-size ($x\longrightarrow 0$) limit. 
In this case, the two variables can couple together to form a dimensionless
finite quantity $\xi$ provided $z<0$ (where $z=-1/(\alpha+1)$ in the case of fragmentation and $\alpha$
is the homogeneity exponent of the breakup kernel).
Recently, we have shown using similar arguments that the fragmentation equation breaks down at $\alpha\leq -1$ \cite{kn.has13}, 
where it is believed to show a shattering transition which is essentially described as the opposite 
phenomena of the gelation transition\cite{kn.has15}.
Returning to the Smoluchowski
equation, it is clear that $z<0$ if $\lambda>1$ and hence the two variables
couple together in the product form to make $\xi$; therefore it cannot be a finite quantity 
in the scaling regime.
We thus argue that the scaling is  
violated for the whole time axis at $\lambda\geq 1$. 

\noindent
Let us check the prototypical case $\lambda=2$ for the product model since this is the best known
model and claimed to exhibit gelation transition. We show that, with this model, one cannot
obtain even one single quantity which is physically meaningful and correct at the same time.
The exact solution for the concentration $\psi(x,t)$ is found to be
$\psi(x,t)\sim t^{x-1}x^{x-2}{\rm exp}[-xt]/x!$ at $t<t_{cr}$ \cite{kn.has8,kn.has11}. 
First, note that $xt$ is the dimensionless quantity which clearly means that $x$ bears the dimension
inverse of time and $z<0$. Trying to look for the dimensional consistency in the rest of the
term of the solution will only yields frustrating results. 
On the other hand, $t_{cr}=1$ for monodisperse 
initial condition. It is too short a time to seek scaling solution since it is the solution
in the asymptotic regime where the initial condition is irrelevant. Therefore, the scaling cannot hold
at $t<1$. Furthermore, one can check and find out that it cannot satisfy Eq. (1),
which is essential for a faithful solution. We show further inconsistencies of this model below.

\noindent
For the monodisperse initial condition, the equation for $M_0(t)$ in Eq.(2) has the 
solution $M_0(t)= 1-t/2$, when $M_1(t)$ is a conserved
quantity. This is believed to be true at $t<1$ if we assume, for arguments sake, that gelation occurs at $t>t_{cr}$.
This shows that the number of particles present in the system decreases in a much slower fashion than the corresponding
non-gelling model,  where it decays 
in the power-law form. Note that, should it actually be the gelling model 
then there would already be an indication of it in the faster decrease of $M_0(t)$  
at least when it is close to $t_{cr}=1$ since at $t>1$, more and more
particles are believed to be lost in the gel phase, which is held responsible for mass violation. 
In addition, the solution for $M_0(t)$ also implies that after a definite time, 
the number density becomes a negative quantity, which is unphysical. In this model, we find that the aggregation rate 
$a(x,t)=xM_1(t)$. This means that if the mass is a conserved quantity then $x$ must bear the dimension
inverse of time and hence $xt$ is the dimensionless quantity.
We already argued that $xt$ cannot stay as a finite quantity in the scaling regime, hence scaling is violated.
On the other hand, if the mass is not assumed to be a 
conserved quantity, but decays as $1/t$, which is 
claimed to be the case at $t>t_{cr}$, then we get $a(x,t)\sim x/t$. 
This is again unphysical since $x$ becomes a dimensionless quantity
and Eq. (1) loses its stochastic nature. This means that Eq. (1) breaks down for the whole time axis.
Solving Eq. $(2)$ for the second moment with monodisperse initial condition,
we obtain $M_2(t)=1/(1-t)$. If the mass is a conserved quantity, then we obtain $w(t)=M_2(t)$ 
and $s(t)=1/M_0(t)$ and hence $s(t)\neq w(t)$ if $z<0$.
This contradicts our previous observation that $s(t)=w(t)$ for all $z$. 
We find that both $s(t)$ and $w(t)$ suffer a singularity but at two different points along the time axis. 
However, the singularity of the simple mean has never been noticed before. Note that below $t_{cr}$, 
the decrease in the number density $M_0(t)$ and the increase in the mean cluster size $w(t)$ are not commensurable.
For example, below but close to $t_{cr}$, the number density decreases very slowly, while
$w(t)$ increases faster than any power-law can predict.

\noindent
Most important of all, it is believed that at $t>1$ the 
mass is no longer a conserved quantity and it decreases as $1/t$. Where is the mass going? 
It has been argued that the 
finite size particle (sol) is lost to the infinite cluster (gel) \cite{kn.has8} and hence gelation transition.
That is, the system contains two different kinds of particles: sol and gel.   
The transition of sol particles 
to the gel particle was held responsible for the violation mass conservation and for the scaling \cite{kn.has8}.
If it is so, then it can only mean that $\psi(x,t)$, 
and hence its moment $M_1(t)$ or $w(t)$, is incapable of taking into account the gel particle despite 
being present in the system.  
Therefore, $M_2(t)$
cannot bear any information about the gel particle, and hence its 
divergence should not be taken as the appearence of 
infinite gel but quite the opposite. This is again self-contradictory. 
It has been argued that at $t\geq t_{cr}$, the concentration 
$\psi(x,t)$ has asymptotically a power-law behaviour $\psi(x,t) \sim x^{-5/2}/t$
\cite{kn.has8}. Firstly, if this is to be the solution of Eq. (1), then it must satisfy  
it in the first place, at least it must remain dimensionally faithful. One can insert this power-law
form of the solution into Eq. (1) and finds that it cannot satisfy it.  Secondly, the solution
implies that $x^{-5/2}/t$ must bear the dimension of the concentration. However, 
one cannot find
any physical basis to support it. Finally, one cannot obtain any physically 
meaningful expression for the moments if $\psi(x,t)$ has the power-law form. 
This holds true in all closely connected problems like the kinetics of fragmentation or the 
random sequential adsorption process \cite{kn.has13}. Note that the aggregation rate 
for both the sum models with $\lambda=2$ contains $M_2(t)$ and so does Eq. (1). 
So, if the divergence of weighted mean $w(t)=M_2(t)$ is considered 
to detect the appearence of gelation, then 
Eq. (1) itself becomes invalid at the singular point. Hence solving Eq. (1) at $t>t_{cr}$
to find $\psi(x,t)$ and the different criterion for gelation i.e. different moments of $\psi(x,t)$
are no longer valid. Note further that $M_2(t)$ diverges only in one particular case 
i.e. at $\lambda=2$ of the product model. 
There do not exist another $\beta$ vis-a-vis a $\lambda$ value 
for which one can show that the second moment
diverges at a finite time. Therefore, one cannot generalise that for $\lambda>1$ the second moment
diverges at a finite time.
It is thus claer that one cannot obtain a self consistent solution in the so called gelling
regime rather one only obtains results which are either unphysical or self-contradictory.

\vspace{3mm}

\noindent
Note that Eq. $(1)$ is a one dimensional mean-field model that ignores
fluctuation, detailed nature of cluster-cluster interaction and their spatial 
correlation and the shape of the aggregating cluster.
It also does not bear any parameter that could explain thermodynamic or
mechanical properties of the system. Instead, it assumes that the system is highly 
diluted so that merging of two clusters into one is not influenced by the presence
of other clusters and they merge into one without failure as soon as they meet. 
Therefore, it is quite natural that as many
other one dimensional mean-field theories in statistical physics it cannot describe phase transition such as sol-gel.
In fact, the sol-gel transition is characterized by a sudden rise in 
viscosity and enthalpy accompanied by the appearance
of non-zero shear modulus near the gel point. This the Smoluchowski's equation can neither explain
quantitatively nor qualitatively. This does not mean that we underestimate the importance of this model.
It still remains one of the few theoretical approaches through which one can obtain
a quantitative comparison of experimental data or data obtained from 
extensive numerical simulation. It is important to mention that the form of the
scaling {\it ansatz} and its exponents obtained from Smoluchowski's equation are 
found to be the general property of all growth phenomena
especially the cluster-cluster aggregation process.

\vspace{3mm}

\noindent
In conclusion, we have given an insightful foundation to the meaning of the scaling ansatz 
of the highly non-trivial Smoluchowski
equation. We hope this will enrich our understanding on the scaling theory. 
We have demonstrated that the Smoluchowski equation cannot describe the gelation transition, instead
it breaks down for all time at $\lambda \geq 1$, which is also accompanied by violation of scaling.
To further support our arguments we have considered the best known gelling model 
and shown that this model cannot 
give any quantity which is physically meaningful and self-consistent.
Finally, we hope this work will be useful in other stochastic processes described by rate equation 
to gain deep insight of the systems especially of the scaling theory.

\vspace{3mm}

\noindent
 MKH is grateful to Dr. G. J. Rodgers for useful correspondence and acknowledges the Alexander von Humboldt Foundation for granting the fellowship.

\end{document}